\documentclass[twocolum]{article}
\usepackage[accepted]{aistats2018}
\pdfoutput=1

\usepackage[utf8]{inputenc} 
\usepackage[T1]{fontenc}    
\usepackage{hyperref}       
\usepackage{url}            
\usepackage{booktabs}       
\usepackage{amsfonts}       
\usepackage{nicefrac}       
\usepackage{microtype}      
\usepackage{amsmath}
\usepackage{amssymb}
\usepackage{mathtools}
\usepackage{bm}
\usepackage{smartdiagram}   
\usesmartdiagramlibrary{additions} 
\usetikzlibrary{arrows}
\usetikzlibrary{arrows, positioning, shadows}
\usepackage{algorithm}
\usepackage{algorithmicx}
\usepackage{algpseudocode}
\usepackage{todonotes}
\usepackage{dsfont}
\usepackage{color}
\usepackage[tight,ugly]{units}
\usepackage{amsthm}
\usepackage[normalem]{ulem}
\usepackage[numbers]{natbib}
\usepackage{subfigure}

\usepackage{enumitem}
\setenumerate{leftmargin=4mm}
\setitemize{leftmargin=4mm}

\algnewcommand{\Initialize}[1]{%
	\State \textbf{initialize:}
}
\newtheorem{theorem}{Theorem}
\theoremstyle{definition}

\newtheorem{lemma}{Lemma}
\theoremstyle{remark}
\newtheorem{remark}{Remark}

\renewcommand{\vec}{\boldsymbol }
\newcommand{\mat}{\boldsymbol}
\newcommand{\aug}[1]{\widetilde{#1}}
\newcommand{\prob}{p}

\newcommand{\modelname}{GP-MPC}

\newcommand{\var}{\mathrm{var}}
\newcommand{\cov}{\mathrm{cov}}
\newcommand{\R}{\mathds{R}}
\newcommand{\E}{\mathds{E}}
\renewcommand{\d}{d}
\newcommand{\Hamiltonian}{\mathcal{H}}
\newcommand{\T}{^T}
\newcommand{\inv}{^{-1}}

\newcommand{\gaussx}[3]{\mathcal{N}\big(#1\,|\,#2,#3\big)}

\definecolor{darkgreen}{rgb}{0,0.6,0}

\newcommand{\blue}[1]{{\color{blue}#1}}
\newcommand{\red}[1]{{\color{red}#1}}

\begin{document}
\runningtitle {Data-Efficient RL with Probabilistic MPC}
\twocolumn[

\aistatstitle{Data-Efficient Reinforcement Learning with \\
Probabilistic Model Predictive Control}
\aistatsauthor{
  Sanket Kamthe    \And
  Marc Peter Deisenroth
  }
\aistatsaddress  { 
Department of Computing\\
Imperial College London  \And
Department of Computing\\
Imperial College London
}
  

  
] 

 \begin{abstract}
   Trial-and-error based reinforcement learning (RL) has seen rapid advancements in recent times, especially with the advent of deep neural networks. However, the majority of autonomous RL algorithms require a large number of interactions with the environment. A large number of interactions may be impractical in many real-world applications, such as robotics, and many practical systems have to obey limitations in the form of state space or control constraints. To reduce the number of system interactions while simultaneously handling constraints, we propose a model-based RL framework based on probabilistic Model Predictive Control (MPC). In particular, we propose to learn a probabilistic transition model using Gaussian Processes (GPs) to incorporate model uncertainty into long-term predictions, thereby, reducing the impact of model errors. We then use MPC to find a control sequence that minimises the expected long-term cost. We provide theoretical guarantees for first-order optimality in the GP-based transition models with deterministic approximate inference for long-term planning. We demonstrate that our approach does not only achieve state-of-the-art data efficiency, but also is a principled way for RL in constrained environments.
 \end{abstract}


 \section{Introduction}
%
%
%
Reinforcement learning (RL) is a principled mathematical framework for experience-based autonomous learning of control policies. Its trial-and-error learning process is one of the most distinguishing features of RL~\citep{Sutton1998}. 
Despite many recent advances in RL~\cite{Mnih2015,Silver2016,Yahya2016}, a  main limitation of current RL algorithms remains its data inefficiency, i.e., the required number of interactions with the environment is impractically high. For example, many RL approaches in problems with low-dimensional state spaces and fairly benign dynamics require thousands of trials to learn. This \emph{data inefficiency} makes learning in real control\slash robotic systems without task-specific priors impractical and prohibits RL approaches in more challenging scenarios.

A promising way to increase the data efficiency of RL without inserting task-specific prior knowledge is to learn models of the underlying system dynamics. When a good model is available, it can be used as a faithful proxy for the real environment, i.e., good policies can be obtained from the model without additional interactions with the real system. However, modelling the underlying transition dynamics accurately is challenging and inevitably leads to model errors. To account for model errors, it has been proposed to use probabilistic models~\cite{Schneider1997,Deisenroth2011c}. By explicitly taking model uncertainty into account, the number of interactions with the real system can be substantially reduced. For example, in~\cite{Deisenroth2011c,Pan2014,Deisenroth2015,Cutler2015}, the authors use Gaussian processes (GPs) to model the dynamics of the underlying system. The PILCO algorithm~\cite{Deisenroth2011c} propagates uncertainty through time for long-term planning and learns parameters of a feedback policy by means of gradient-based policy search. It achieves an unprecedented data efficiency for learning control policies for from scratch. 

While the PILCO algorithm is data efficient, it has few shortcomings: 1) Learning closed-loop feedback policies needs the full planning horizon to stabilise the system, which results in a significant computational burden; 2) It requires us to specify a parametrised policy a priori, often with hundreds of parameters; 3) It cannot handle state constraints; 4) Control constraints are enforced by using a differentiable squashing function that is applied to the RBF policy. This allows PILCO to explicitly take control constraints into account during planning.  However, this kind of constraint handling can produce unreliable predictions near constraint boundaries~\cite{Schattler2012, Naidu2003, Mayne2000}.

In this paper, we develop an RL algorithm that is a) data efficient, b) does not require to look at the full planning horizon, c) handles constraints naturally, d) does not require a parametrised policy, e) is theoretically justified. The key idea is to reformulate the optimal control problem with learned GP models as an equivalent deterministic problem, an idea similar to~\cite{Ng2000}. This reformulation allows us to exploit Pontryagin's maximum principle to find optimal control signals while handling constraints in a principled way. We propose probabilistic model predictive control (MPC) with learned GP models, while propagating uncertainty through time. The MPC formulation allows to plan ahead for relatively short horizons, which limits the computational burden and allows for infinite-horizon control applications. 
Our approach can find optimal trajectories in constrained settings, offers an increased robustness to model errors and an unprecedented data efficiency compared to the state of the art.

\paragraph{Related Work}
\textit{Model-based RL:}
A recent survey of model based RL in robotics~\citep{Polydoros2017} highlights the importance of models for building adaptable robots. Instead of GP dynamics model with a zero prior mean (as used in this paper) an RBF network and linear mean functions are proposed~\citep{Cutler2015,Bischoff2014}. This accelerates learning and facilitates transferring a learned model from simulation to a real robot. Even implicit model learning can be beneficial: The UNREAL learner proposed in~\citep{Jaderberg2016} learns a predictive model for the environment as an auxiliary task, which helps learning. 

\textit{MPC with GP transition models:}
GP-based predictive control was used for boiler  and building control~\cite{Grancharova2008,Nghiem2017}, but the model uncertainty was discarded. In~\cite{Klenske2016}, the predictive variances were used within a GP-MPC scheme to actively reject periodic disturbances, although not in an RL setting. Similarly, in~\citep{Ostafew2015,Ostafew2016}, the authors used a GP prior to model additive noise and model is improved episodically. In~\cite{Boedecker2014, Nghiem2017}, the authors considered MPC problems with GP models, where only the GP's posterior mean was used while ignoring the variance for planning. 
MPC methods with deterministic models are useful only when model errors and system noise can be neglected in the problem~\citep{Kocijan2016,Girard2003}.

\textit{Optimal Control:}
The application of optimal control theory for the models based on GP dynamics employs some structure in the transition model, i.e., there is an explicit assumption of control affinity~\citep{Hennig2011, Pan2014, Pan2015, Boedecker2014} and linearisation via locally quadratic approximations~\citep{Boedecker2014, Pan2014}. The AICO model~\citep{Toussaint2009} uses approximate inference with (known) locally linear models. The probabilistic trajectories for model-free RL in~\citep{Rawlik2012} are obtained by reformulating the stochastic optimal control problem as KL divergence minimisation. We implicitly linearise the transition dynamic via moment matching approximation.

\paragraph{Contribution}
The contributions of this paper are the following: 1) We propose a new `deterministic' formulation for probabilistic MPC with learned GP models and uncertainty propagation for long-term planning. 
2) This reformulation allows us to apply Pontryagin's Maximum Principle (PMP) for the open-loop planning stage of probabilistic MPC with GPs. Using the PMP we can handle control constraints in a principled fashion while still maintaining necessary conditions for optimality. 
3) The proposed algorithm is not only theoretically justified by optimal control theory, but also achieves a state-of-the-art data efficiency in RL while maintaining the probabilistic formulation.
4) Our method can handle state and control constraints while preserving its data efficiency and optimality properties.

 \section{Controller Learning via Probabilistic MPC}
We consider a stochastic dynamical system with states $\vec x\in\R^D$ and admissible controls (actions) $\vec u\in\mathcal U\subset \R^U$, where the state follows Markovian dynamics
\begin{align}
\vec x_{t+1} = f(\vec x_t, \vec u_t) + \vec w
\end{align}
with an (unknown) transition function $f$ and i.i.d. system noise $\vec w\sim\mathcal N(\vec 0,\mat Q)$, where $\mat Q = \text{diag}(\sigma_1^2, \dotsc, \sigma_D^2)$. In this paper, we consider an RL setting where we seek control signals $\vec u_0^*, \dotsc, \vec u_{T-1}^*$ that minimise the expected long-term cost
\begin{align}
J = \E[\Phi(\vec x_T)] + \sum\nolimits_{t=0}^{T-1} \E[\ell(\vec x_t, \vec u_t)]\,,
\label{eq:cost-to-go}
\end{align}
where $\Phi(\vec x_T)$ is a terminal cost and $\ell(\vec x_t, \vec u_t)$ the stage cost associated with applying control $\vec u_t$ in state $\vec x_t$.
 We assume that the initial state is Gaussian distributed, i.e., $ p(\vec x_0) = \mathcal N(\vec\mu_0, \mat\Sigma_0)$. 

For data efficiency, we follow a model-based RL strategy, i.e., we learn a model of the unknown transition function $f$, which we then use to find open-loop\footnote{`Open-loop' refers to the fact that the control signals are independent of the state, i.e., there is no state feedback incorporated.} optimal controls $\vec u_0^*,\dotsc, \vec u_{T-1}^*$ that minimise~\eqref{eq:cost-to-go}. After every application of the control sequence, we update the learned model with the newly acquired experience and re-plan. Section~\ref{sec:model learning} summarises the model learning step; Section~\ref{sec:open loop learning} details how to obtain the desired open-loop trajectory.

\subsection{Probabilistic Transition Model}
\label{sec:model learning}
We learn a probabilistic model of the unknown underlying dynamics $f$ to be robust to model errors~\cite{Schneider1997, Deisenroth2011c}. In particular, we use a Gaussian process (GP) as a prior $p(f)$ over plausible transition functions $f$. 

A GP is a probabilistic non-parametric model for regression. In a GP, any finite number of function values is jointly Gaussian distributed~\cite{Rasmussen2006}. A GP is fully specified by a mean function $m(\cdot)$ and a covariance function (kernel) $k(\cdot, \cdot)$. 

The inputs for the dynamics GP are given by tuples $\aug{\vec x}_t:=(\vec x_t, \vec u_t)$, and the corresponding targets are $\vec x_{t+1}$.  We denote the collections of training inputs and targets by $\aug{\mat X}, \vec y$, respectively. Furthermore, we assume a Gaussian (RBF, squared exponential) covariance function
\begin{align}
\hspace{-2mm}k(\aug{\vec x}_i, \aug{\vec x}_j) = \sigma_f^2 \exp\left(-\tfrac{1}{2}(\aug{\vec x}_i - \aug{\vec x}_j)^T\mat L^{-1}(\aug{\vec x}_i - \aug{\vec x}_j)\right)\,,
\label{eq:gaussian kernel}
\end{align}
where $\sigma_f^2$ is the signal variance and $\mat L = \text{diag}(l_1,\dotsc,l_{D+U})$ is a diagonal matrix of length-scales $l_1, \dotsc, l_{D+U}$. 
The GP is trained via the standard procedure of evidence maximisation~\cite{MacKay1998,Rasmussen2006}. 

We make the standard assumption that the GPs for each target dimension of the transition function $f:\R^D\times \mathcal U\to\R^D$ are independent. 
For given hyper-parameters, training inputs $\aug{\mat X}$, training targets $\vec y$ and a new test input $\aug{\vec x}_*$, the GP yields the predictive distribution $p(f(\aug{\vec x}_*)|\aug{\mat X}, \vec y) = \mathcal N(f(\aug{\vec x}_*)|m(\aug{\vec x}_*), \Sigma(\aug{\vec x}_*))$, where 
\begin{align}
&m(\aug{\vec x}_*) = [m_1(\aug{\vec x}_*), \dotsc, m_D(\aug{\vec x}_*)]\T \\
 &m_d(\aug{\vec x}_*) = k_d(\aug{\vec x}_*, \aug{\mat  X})(\mat K_d + \sigma_d^2\mat I)^{-1}\vec y_d\\
&\mat \Sigma(\aug{\vec x}_*) = \text{diag} \big(\sigma_1^2(\aug{\vec x}_*), \dotsc, \sigma_D^2(\aug{\vec x}_*)\big) \\
& \sigma_d^2  = \sigma_{f_d}^2 - k_d(\aug{\vec x}_*, \aug{\mat X})(\mat K_d+\sigma_d^2\mat I)^{-1}k_d(\aug{\mat X}, \aug{\vec x}_*)\,,
\end{align}
for all predictive dimensions $d = 1,\dotsc, D$. 

\subsection{Open-Loop Control}
\label{sec:open loop learning}
To find the desired open-loop control sequence $\vec u_0^*, \dotsc, \vec u_{T-1}^*$, we follow a two-step procedure proposed in~\cite{Deisenroth2011c}. 1) Use the learned GP model to predict the long-term evolution $p(\vec x_1), \dotsc, \vec p(\vec x_T)$ of the state for a given control sequence $\vec u_0, \dotsc, \vec u_{T-1}$. 2) Compute the corresponding expected long-term cost~\eqref{eq:cost-to-go} and find an open-loop control sequence $\vec u_0^*, \dotsc, \vec u_{T-1}^*$ that minimises the expected long-term cost. In the following, we will detail these steps.

\subsubsection{Long-term Predictions}
\label{sec:long-term predictions}
To obtain the state distributions $p(\vec x_1), \dotsc, p(\vec x_T)$ for a given control sequence $\vec u_0, \dotsc, \vec u_{T-1}$, we iteratively predict 
\begin{align}
\label{eq:MM uncertainty propagation}
p(\vec x_{t+1}|\vec u_t) = \iint p(\vec x_{t+1}|\vec x_t, \vec u_t) p(\vec x_t)p(f)df d\vec x_t
\end{align}
for  $t = 0, \dotsc, T-1\,,$ by making a \emph{deterministic} Gaussian approximation to $p(\vec x_{t+1}|\vec u_t)$ using moment matching~\cite{Girard2003,Quinonero-Candela2005a,Deisenroth2011c}. This approximation has been shown to work well in practice in RL contexts~\cite{Deisenroth2011c, Deisenroth2015, Cutler2015, Bischoff2013, Bischoff2014, Pan2014, Pan2015} and can be computed in closed form when using the Gaussian kernel~\eqref{eq:gaussian kernel}.

A key property that we exploit is that moment matching allows us to formulate the uncertainty propagation in~\eqref{eq:MM uncertainty propagation} as  a `deterministic system function'
\begin{align}
\label{eqn:Augment_Defn1}
\vec z_{t+1} = f_{MM}(\vec z_t, \vec u_t)\,, \quad \vec z_t := [\vec \mu_t, \mat\Sigma_t]\,,
\end{align}
where $\vec\mu_t, \mat\Sigma_t$ are the mean and the covariance of $p(\vec x_t)$. For a deterministic control signal $\vec u_t$ we further define the moments of the control-augmented distribution $p(\vec x_t, \vec u_t)$ as
\begin{align}
\hspace{-3mm}\aug{\vec z}_t := [\aug{\vec \mu}_t, \aug{\vec \Sigma}_t],\,
\aug{\vec \mu}_t = 
[
\vec\mu_t
,
\vec u_t
]
,\,
\aug{\vec \Sigma}_t= 
\text{blkdiag}[
\mat\Sigma_t ,\, \mat 0
]
\,,
\label{eq:control-augmented distribution}
\end{align}
such that~\eqref{eqn:Augment_Defn1} can equivalently be written as the deterministic system equation
\begin{align}
\label{eqn:Augment_Defn}
\vec z_{t+1} = f_{MM}(\aug{\vec z}_t)\,.
\end{align}

\subsubsection{Optimal Open-Loop Control Sequence}
To find the optimal open-loop sequence $\vec u_0^*, \dotsc, \vec u_{T-1}^*$, we first compute the expected long-term cost $J$ in~\eqref{eq:cost-to-go} using the Gaussian approximations $p(\vec x_1), \dotsc, p(\vec x_T)$ obtained via~\eqref{eq:MM uncertainty propagation} for a given open-loop control sequence $\vec u_0, \dotsc, \vec u_{T-1}$. Second, we find a control sequence that minimises the expected long-term cost~\eqref{eq:cost-to-go}. In the following, we detail these steps.


\paragraph{Computing the Expected Long-Term Cost}
To compute the expected long-term cost in~\eqref{eq:cost-to-go}, we sum up the expected immediate costs
\begin{align}
\E [\ell(\vec x_t, \vec u_t)] = \int \ell(\aug{\vec x}_t) \mathcal N(\aug{\vec x}_t|\aug{\vec\mu}_t, \aug{\mat\Sigma}_t)d\aug{\vec x}_t
\end{align}
for  $t = 0, \dotsc, T-1$. We choose $\ell$, such that this expectation and the partial derivatives $\partial\E[\ell (\vec x_t, \vec u_t)]/\partial\vec x_t$, $\partial\E[\ell (\vec x_t, \vec u_t)]/\partial\vec u_t$ can be computed analytically.\footnote{Choices for $\ell$ include the standard quadratic (polynomial) cost, but also costs expressed as Fourier series expansions or radial basis function networks with Gaussian basis function.} 

Similar to~\eqref{eqn:Augment_Defn1}, this  allows us to define deterministic mappings 
\begin{align}
&\ell_{MM}(\vec z_t, \vec u_t) = \ell_{MM}(\aug{\vec z}_t) := \E[\ell(\vec x_t, \vec u_t)]\\
&\Phi_{MM}(\vec z_T):=\E[\Phi(\vec x_T)]
\end{align}
that map the mean and covariance of $\aug{\vec x}$ onto the corresponding expected costs in~\eqref{eq:cost-to-go}.

\begin{remark}
The open-loop optimisation turns out to be sparse~\cite{Bock1984}. However, optimisation via the value function or dynamic programming is valid only for unconstrained controls. To address this practical shortcoming,  we define Pontryagin's Maximum Principle that allows us to formulate the constrained problem while maintaining the sparsity. We detail this sparse structure for the constrained GP dynamics problem in section~\ref{sec:justification}.
\end{remark}

\subsection{Feedback Control with MPC}
Thus far, we presented a way for efficiently determining an open-loop controller. However, an open-loop controller cannot stabilise the system~\citep{Mayne2000}. Therefore, it is essential to obtain a feedback controller. MPC is a practical framework for this~\citep{Mayne2000,Grune2011}. While interacting with the system MPC determines an $H$-step open-loop control trajectory $\vec u_0^*, \dotsc, \vec u_{H-1}^*$, starting from the current state $\vec x_t$\footnote{A state distribution $p(\vec x_t)$ would work equivalently in our framework.}. Only the first control signal $\vec u_0^*$ is applied to the system. When the system transitions to $\vec x_{t+1}$, we update the GP model with the newly available information, and MPC re-plans $\vec u_0^*, \dotsc, \vec u_{H-1}^*$. This procedure turns an open-loop controller into an implicit closed-loop (feedback) controller by repeated re-planning $H$ steps ahead from the current state. Typically,  $H\ll T$, and MPC even allows for $T=\infty$.

In this section, we provided an algorithmic framework for probabilistic MPC with learned GP models for the underlying system dynamics, where we explicitly use the GP's uncertainty for long-term predictions~\eqref{eq:MM uncertainty propagation}. In the following section, we will justify this using optimal control theory. Additionally, we will discuss how to account for constrained control signals in a principled way without the necessity to warp\slash squash control signals as in~\cite{Deisenroth2011c}.

 \section{Theoretical Justification}
\label{sec:justification}




Bellman's optimality principle~\citep{Bellman1957} yields a recursive formulation for calculating the total expected cost~\eqref{eq:cost-to-go} and gives a sufficient optimality condition. PMP~\citep{Pontryagin1962} provides the corresponding necessary optimality condition. PMP allows us to compute gradients $\partial J/\partial \vec u_t$ of the expected long-term cost w.r.t. the variables that only depend on variables with neighbouring time index, i.e., $\partial J/\partial \vec u_t$ depends only variables with index $t$ and $t+1$.  Furthermore, it allows us to explicitly deal with constraints on states and controls. In the following, we detail how to solve the optimal control problem (OCP) with PMP for learned GP dynamics and deterministic uncertainty propagation. We additionally provide a computationally efficient way to compute derivatives based on the maximum principle.

 

To facilitate our discussion we first define some notation. Practical control signals are often constrained. We formally define a class of \textit{admissible controls} $ \mathcal U $ that are piecewise continuous functions defined on a compact space $ U\subset \R^U $. This definition is fairly general, and commonly used zero-order-hold or first-order-hold signals satisfy this requirement.  Applying admissible controls to the deterministic system dynamics $f_{MM}$ defined in~\eqref{eqn:Augment_Defn} yields a set $ \mathcal{Z} $ of 
\textit{admissible controlled trajectories}.  We define the tuple $ (\mathcal{Z}, f_{MM}, \mathcal{U}) $ as our control system. For a single admissible control trajectory $\vec u_{0:H-1}$, there will be a unique trajectory $\vec z_{0:H}$, and the pair  $(\vec z_{0:H}, \vec u_{0:H-1})$ is called an admissible controlled trajectory~\citep{Schattler2012}.

We now define the control-Hamiltonian~\citep{Clarke1990,Schattler2012,Todorov2009} for this control system as  
	\begin{align}
    \label{defn:hamiltonian}
	\Hamiltonian(\vec \lambda_{t+1}, \vec z_t, \vec u_t) =  \ell_{MM}(\vec z_t, \vec u_t) + \vec \lambda_{t+1}^T f_{MM}(\vec z_t, \vec u_t) .
	\end{align}
This formulation of the control-Hamiltonian is the centre piece of the Pontryagin's approach to the OCP. 
The vector $ \vec \lambda_{t+1} $ can be viewed as a Lagrange multiplier for dynamics constraints associated with the OCP~\cite{Clarke1990,Schattler2012}.  

To successfully apply PMP we need the system dynamics to have a unique solution for a given control sequence. Traditionally, this is interpreted as the system is `deterministic'.  This interpretation has been considered a limitation of PMP~\citep{Todorov2009}. In this paper, however, we exploit the fact that the moment-matching approximation~\eqref{eq:MM uncertainty propagation} is a deterministic operator, similar to the projection used in EP~\cite{Opper2001, Minka2001}. This yields the `deterministic' system equations~\eqref{eqn:Augment_Defn1},~\eqref{eqn:Augment_Defn} that map moments of the state distribution at time $t$ to moments of the state distribution at time $t+1$.

\subsection{Existence/Uniqueness of a Local Solution}
To apply the PMP we need to extend some of the important characteristics of ODEs to our system.  In particular, we need to show the existence and uniqueness of a (local) solution to our difference equation~\eqref{eqn:Augment_Defn}.

For existence of a solution we need to satisfy the difference equation point-wise over the entire horizon and for uniqueness we need the system to have only one singularity. 
For our discrete-time system equation (via the moment-matching approximation) in~\eqref{eqn:Augment_Defn1} we have the following
\begin{lemma}
\label{lemma:fmm L-continuous}
	The moment matching mapping $ f_{MM} $ is Lipschitz continuous for controls defined over a compact set $\mathcal{U}$. 
\end{lemma}
The proof is based on bounding the gradient of $f_{MM}$ and detailed in the supplementary material.
%
Existence and uniqueness of the trajectories for the moment matching difference equation are given by 
\begin{lemma}
\label{lemma:unique}
	A solution of $ \vec z_{t+1} = f_{MM}(\vec z_t, \vec u_t) $ exists and is unique. 
\end{lemma}
\paragraph{Proof Sketch} 
Difference equations always yield an answer for a given input. Therefore, a solution trivially exists. Uniqueness directly follows from the Picard-Lindel\"{o}f theorem, which we can apply due to Lemma~\ref{lemma:fmm L-continuous}. This theorem requires the discrete-time system function to be deterministic~(see Appendix B of~\citep{Schattler2012}). Due to our re-formulation of the system dynamics~\eqref{eqn:Augment_Defn1}, this follows directly, such that the $ \vec z_{1:T} $ for a given control sequence $ \vec u_{0:T-1} $ are unique.



\subsection{Pontryagin's Maximum Principle for GP Dynamics}
With Lemmas~\ref{lemma:fmm L-continuous}~\ref{lemma:unique} and the definition of the control-Hamiltonian~\ref{defn:hamiltonian} we can now state PMP for the control system $ (\vec{Z}, f_{MM}, \mathcal{U}) $ as follows:
\begin{theorem}
	Let $ (\vec z_t^*, \vec u_t^*) $, $ 0 \le t \le H-1 $ be an admissible controlled trajectory defined over the horizon $H$. If $ (\vec z_{0:H}^*, \vec u_{0:H-1}^*)$ is optimal,
	then there exists an ad-joint vector $ \vec \lambda_{t} \in \R^D\setminus\{\vec 0\} $ 
	 satisfying the following conditions:
	\begin{enumerate}
		\item
		 Ad-joint equation: The ad-joint vector $\vec \lambda_t $ is a solution to the discrete difference equation
		\begin{align}
		\label{eqn:adjoint}
		\hspace{-4mm}\vec \lambda_{t}^T &=  \dfrac{\partial}{\partial \vec z_t}\ell_{MM}(\vec z_t, \vec u_t) + \vec\lambda^{T}_{t+1}\frac{\partial f_{MM}(\vec z_t, \vec u_t)}{\partial \vec z_t}\,.
		\end{align}
        \item	
		Transversality condition: At the endpoint the ad-joint vector  $ \vec \lambda_{H} $ satisfies
		\begin{align}
        \label{eq:terminal_condn}
		\vec \lambda_{H}  &= \frac{\partial}{\partial \vec z_H}\Phi_{MM}(\vec z_H)\,.
        		\end{align}
\item		
		Minimum Condition: For $t = 0 , \dotsc, H-1$, we have
		\begin{align}
		\label{eqn:Hamiltonian_min}
	\hspace{-4mm}	\Hamiltonian ( \vec \lambda_{t+1}, \vec z_t^*, \vec u_t^*) &= \min_{\vec \nu}\Hamiltonian(\vec \lambda_{t+1}, \vec z_t^*, \vec \nu)
		\end{align}
        for all $\vec \nu \in\mathcal U$.
	\end{enumerate} 
\end{theorem}


\begin{remark}
The minimum condition~\eqref{eqn:Hamiltonian_min} can be used to find an optimal control. The Hamiltonian is minimised point-wise over the admissible control set $ \mathcal{U} $: For every $ t =0,\dotsc, H-1 $ we find optimal controls  $\vec u_t^*\in\arg\min_{\vec\nu}  \Hamiltonian(\vec \lambda_{t+1}, \vec z_t^*, \vec\nu) $.  The minimisation problem possesses additional variables $ \vec \lambda_{t+1} $. These variables can be interpreted as Lagrange multipliers for the optimisation. They capture the impact of the control $\vec u_t$ over the whole trajectory and, hence, these variables make the optimisation problem sparse~\citep{Diehl2014}. For the GP dynamics we compute the multipliers $\vec \lambda_t $ in closed form, thereby, significantly reducing the computational burden to minimise the expected long-term cost $ J $ in~\eqref{eq:cost-to-go}. 
We detail this calculation in section~\ref{sec:efficient_grad}.
\end{remark}
\begin{remark}
In the optimal control problem, we aim to find an admissible control trajectory that minimizes the cost subject to possibly additional constraints. PMP gives first-order optimality conditions over these admissible controlled trajectories and can be generalised to handle additional state and control constraints ~\citep{Todorov2009,Naidu2003,Schattler2012}.
\end{remark}
\begin{remark}
The Hamiltonian $\Hamiltonian$ in~\eqref{eqn:Hamiltonian_min} is constant for unconstrained controls in time-invariant dynamics and equals $0$ everywhere when the final time $H$ is not fixed~\citep{Schattler2012}.
\end{remark}
\begin{remark}
For linear dynamics the proposed method is a generalisation of iLQG~\citep{Todorov2005}: The moment matching transition $f_{MM}$ implicitly linearises the transition dynamics at each time step, whereas in iLQG an explicit local linear approximation is made. For a linear $f_{MM}$ and a quadratic cost $\ell$ we can write the LQG case as shown in Theorem 1 in~\citep{Toussaint2009}. If we iterate with successive corrections to the linear approximations we obtain iLQG. 
\end{remark}

\subsection{Efficient Gradient Computation}
\label{sec:efficient_grad}
With the definition of the Hamiltonian $\Hamiltonian$ in~\eqref{defn:hamiltonian}  we can efficiently calculate the gradient of the expected total cost $J$. For a time horizon $H$
we can write the accumulated cost as the Bellman recursion~\citep{Bellman1957}
\begin{align}
\label{eqn:totalCostRecur}
J_H(\vec z_H) &:= \Phi_{MM}(\vec z_H),  \\
  J_t(\vec z_t) & := \ell_{MM} (\vec z_t, \vec u_t) + J_{t+1}(\vec z_{t+1})
\end{align}
for $t= H-1, \dotsc, 0$.
Since the (open-loop) control $\vec u_t$ only impacts the future costs via $ \vec z_{t+1}= f_{MM}(\vec z_t, \vec u_t)  $ the derivative of the total cost with $\vec u_t$ is given by
\begin{align}
\frac{\partial J_t }{\partial\vec  u_t}  &= \frac{\partial \ell_{MM} (\vec z_t, \vec u_t)}{\partial\vec  u_t} + \frac{\partial J_{t+1} }{\partial\vec  z_{t+1}}  \frac{\partial f_{MM}(\vec z_t, \vec u_t)  }{\partial\vec  u_{t}}\,.
\end{align}
Comparing this expression with the definition of the Hamiltonian~\eqref{defn:hamiltonian}, we see that if we make the substitution $ \vec \lambda_{t+1}^T = \frac{\partial J_{t+1} }{\partial\vec z_{t+1} } $  we obtain 
\begin{align}
\frac{\partial J_t }{\partial \vec u_t} =\frac{\partial \ell_{MM} (\vec z_t, \vec  u_t)}{\partial \vec u_t} + \vec \lambda_{t+1} ^T  \frac{\partial f_{MM}(\vec z_t,\vec  u_t)  }{\partial \vec u_{t}}  = \frac{\Hamiltonian}{\partial\vec  u_t}.
\label{eq:dHdu}
\end{align}
This implies that the gradient of the expected long-term cost w.r.t. $\vec u_t$ can be efficiently computed using the Hamiltonian~\citep{Todorov2009}.
Next we show that the substitution $ \vec \lambda_{t+1}^T = \frac{\partial J_{t+1} }{\partial\vec z_{t+1} } $ is valid for the entire horizon $H$. For the terminal cost $\Phi_{MM}(\vec z_H) $ this is valid by the transversality condition~\eqref{eq:terminal_condn}. For other time steps we differentiate~\eqref{eqn:totalCostRecur} w.r.t. $ \vec z_t $, which yields
\begin{align}
\frac{\partial J_{t} }{\partial \vec z_{t}} &=\frac{\partial \ell_{MM} (\vec z_t,\vec  u_t)}{\partial \vec z_t} + \frac{\partial J_{t+1}}{\partial\vec  z_{t+1}}\frac{\partial\vec  z_{t+1}}{\partial \vec z_{t}}  \\ &=\frac{\partial \ell_{MM} (\vec z_t, \vec u_t)}{\partial\vec  z_t} + \vec \lambda_{t+1}^T\frac{\partial f_{MM}(\vec z_t, \vec u_t) }{\partial \vec z_{t}}\,,
\end{align}
which is identical to the ad-joint equation~\eqref{eqn:adjoint}.  
Hence, in our setting, PMP implies that \emph{gradient descent on the Hamiltonian $ \Hamiltonian $ is equivalent to gradient descent on the total cost}~\eqref{eq:cost-to-go}~\citep{Schattler2012,Diehl2014}. 


Algorithmically, in an RL setting, we find the optimal control sequence $\vec u_0^*, \dotsc, \vec u_{H-1}^*$ as follows: 
\begin{enumerate}
\item For a given initial (random) control sequence $\vec  u_{0:H-1} $ we follow the steps described in section~\ref{sec:long-term predictions} to determine the corresponding trajectory $\vec  z_{1:H} $. Additionally, we compute Lagrange multipliers $ \vec \lambda_{t+1}^T = \frac{\partial J_{t+1} }{\partial \vec z_{t+1} } $ during the forward propagation. Note that traditionally ad-joint equations are propagated backward to find the multipliers~\citep{Clarke1990,Schattler2012}. 
\item Given $\vec \lambda_t $ and a cost function $\ell_{MM}$ we can determine the Hamiltonians $\Hamiltonian_{1:H}$. Then we find a  new control sequence $ \vec u^*_{0:H-1} $ via any gradient descent method using~\eqref{eq:dHdu}. 
\item Return to 1 or exit when converged.   
\end{enumerate}

We use Sequential Quadratic Programming (SQP) with BFGS for Hessian updates~\citep{Nocedal2006}. The Lagrangian of SQP is a partially separable function~\citep{Griewank1982}. In the PMP, this separation is explicit via the Hamiltonians, i.e., the $\Hamiltonian_t$ is a function of variables with index $t$ or $t+1$.   
This leads to a block-diagonal Hessian of SQP Lagrangian~\citep{Griewank1982}. The structure can be exploited to approximate the Hessian via block-updates within BFGS~\citep{Griewank1982, Bock1984} 


 \section{Experimental Results}

We evaluate the quality of our algorithm in two ways: First, we assess whether probabilistic MPC leads to faster learning compared with PILCO, the current state of the art in terms of data efficiency. Second, we assess the impact of state constraints while performing the same task. 


We consider two RL benchmark problems: the under-actuated cart-pole-swing-up and the fully actuated double-pendulum swing-up. In both tasks, PILCO is the most data-efficient RL algorithm to date~\citep{Deisenroth2015}.

\begin{figure}
	\centering
	\subfigure[Cart-pole with constraint.]{
		\includegraphics[height = 3cm]{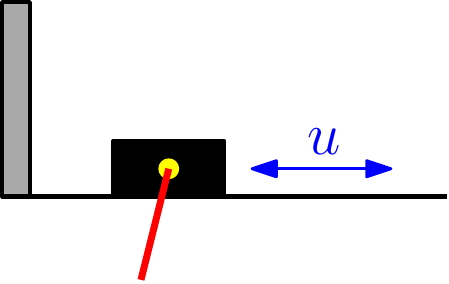}
		\label{fig:cp constraint}
	}
	\hfill
	\subfigure[Double pendulum with constraint.]{
		\includegraphics[height = 3cm]{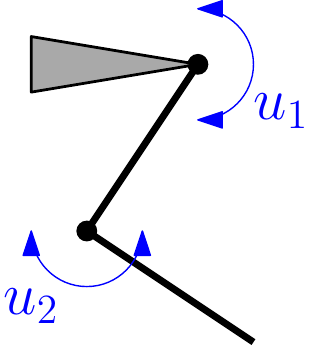}
		\label{fig:dp constraint}
	}
	\caption{State constraints in RL benchmarks. \subref{fig:cp constraint} The position of the cart is constrained on the left side by a wall. \subref{fig:dp constraint} The angle of the inner pendulum cannot enter the grey region.}
\end{figure}

\paragraph{Under-actuated Cart-Pole Swing-Up}
The cart pole system is an under-actuated system with a freely swinging pendulum of $\unit[50]{cm}$ mounted on a cart. The swing-up and balancing task cannot be solved using a linear model~\citep{Raiko2009}.  The cart-pole system state space consists of the position of the cart $x$, cart velocity $\dot{x}$, the angle $\theta$ of the pendulum and the angular velocity $\dot{\theta}$. A horizontal force $u \in [-10, 10]$\,N can be applied to the cart. Starting in a position where the pendulum hangs downwards, the objective is to automatically learn a controller that swings the pendulum up and balances it in the inverted position in the middle of the track.  
\paragraph{Constrained Cart-Pole Swing-Up}
For the state-space constraint experiment we place a wall on the track near the target, see Fig.~\ref{fig:cp constraint}. The wall is at \unit[-70]{cm}, which, along with force limitations, requires the system to swing from the right side. 
 
\paragraph{Fully-actuated Double-Pendulum}
The double pendulum system is a two-link robot arm (links lengths: $\unit[1]{m}$) with two actuators at each joint. The state space consists of 2 angles and 2 angular velocities $[\theta_1, \theta_2, \dot{\theta}_1, \dot{\theta}_2]$~\citep{Deisenroth2015}. 
The torques $u_1$ and $u_2$ are limited to $[ -2, 2]$\,Nm. Starting from a position where both links are in a downwards position, the objective is to learn a control strategy that swings the double-pendulum up and balances it in the inverted position.

\paragraph{Constrained Double-Pendulum}
The double-pendulum has a constraint on the angle of the inner pendulum, so that it only has a $340^\circ$ motion range, i.e., it cannot spin through, see Fig.~\ref{fig:dp constraint}. The constraint blocks all clockwise swing-ups. The system is underpowered, and it has to swing clockwise first for a counter-clockwise swing-up without violating the constraints.

\paragraph{Trials} The general setting is as follows: All RL algorithms start off with a single short random trajectory, which is used for learning the dynamics model. As in~\citep{Deisenroth2011c, Deisenroth2015} the GP is used to predict state differences $\vec x_{t+1}-\vec x_t$. The learned GP dynamics model is then used to determine a controller based on iterated moment matching~\eqref{eq:MM uncertainty propagation}, which is then applied to the system, starting from $\vec x_0 \sim p(\vec x_0)$. Model learning, controller learning and application of the controller to the system constitute a `trial'. After each trial, the hyper-parameters of the model are updated with the newly acquired experience and learning continues.

\paragraph{Baselines}
We compare our \modelname~approach with the following baselines: the PILCO algorithm~\citep{Deisenroth2011c, Deisenroth2015} and a zero-variance GP-MPC algorithm (in the flavour of~\citep{Nghiem2017,Boedecker2014}) for RL, where the GP's predictive variances are discarded. Due to the lack of exploration, such a zero-variance approach within PILCO (a policy search method) does not learn anything useful as already demonstrated in~\citep{Deisenroth2015}, and we do not include this baseline.

We average over 10 independent experiments, where every algorithm is initialised with the same first (random) trajectory. The performance differences of the RL algorithms are therefore due to different approaches to controller learning and the induced exploration.

\subsection{Data Efficiency}
\begin{figure}[tb]
	\centering
    \subfigure[Under-actuated cart-pole swing-up.]{
	\includegraphics[height=3.8cm]{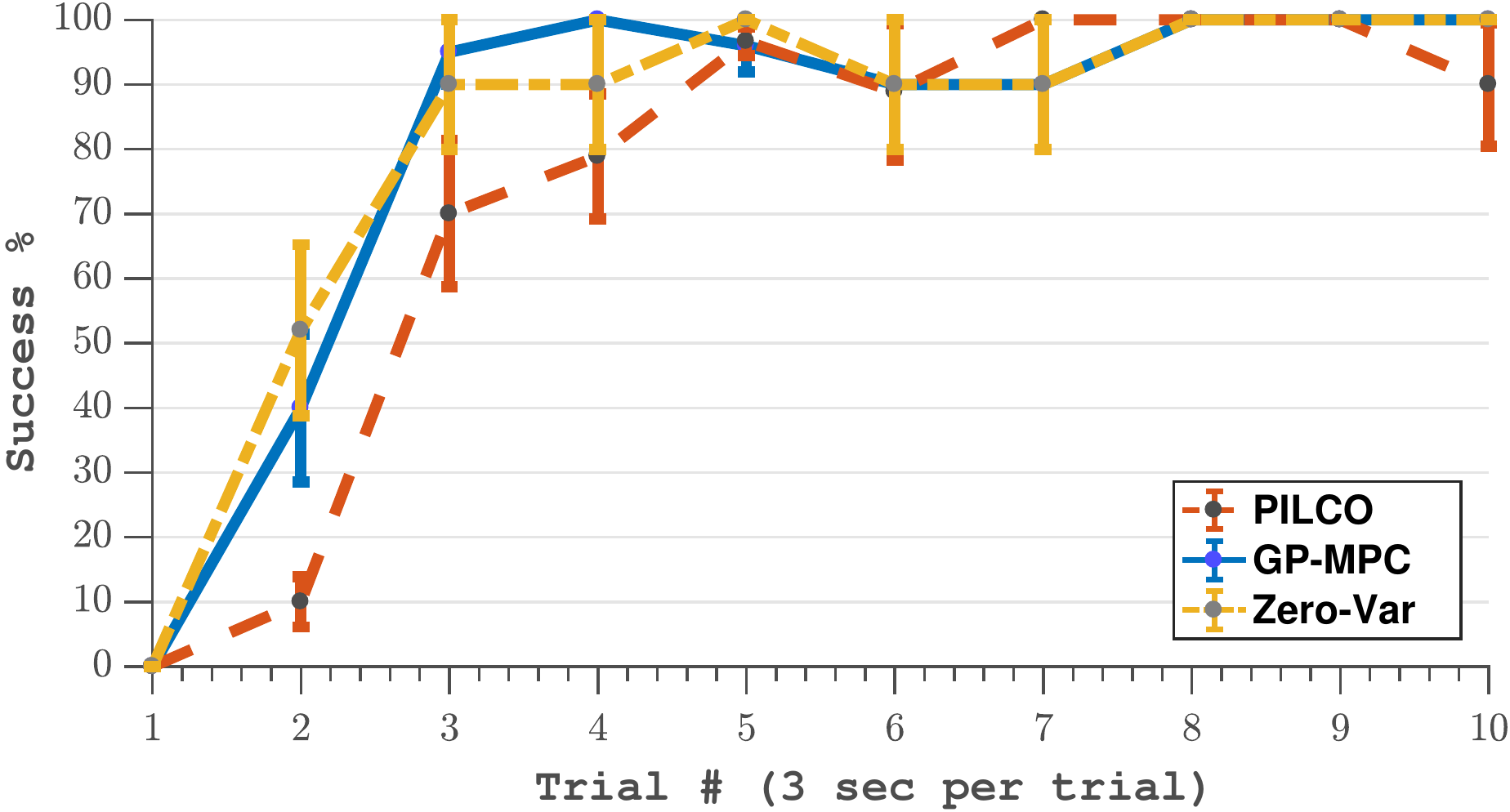}
    \label{fig:cartpole}
    }
    \hfill
    \subfigure[Fully-actuated double-pendulum swing-up.]{
	\includegraphics[height=3.8cm]{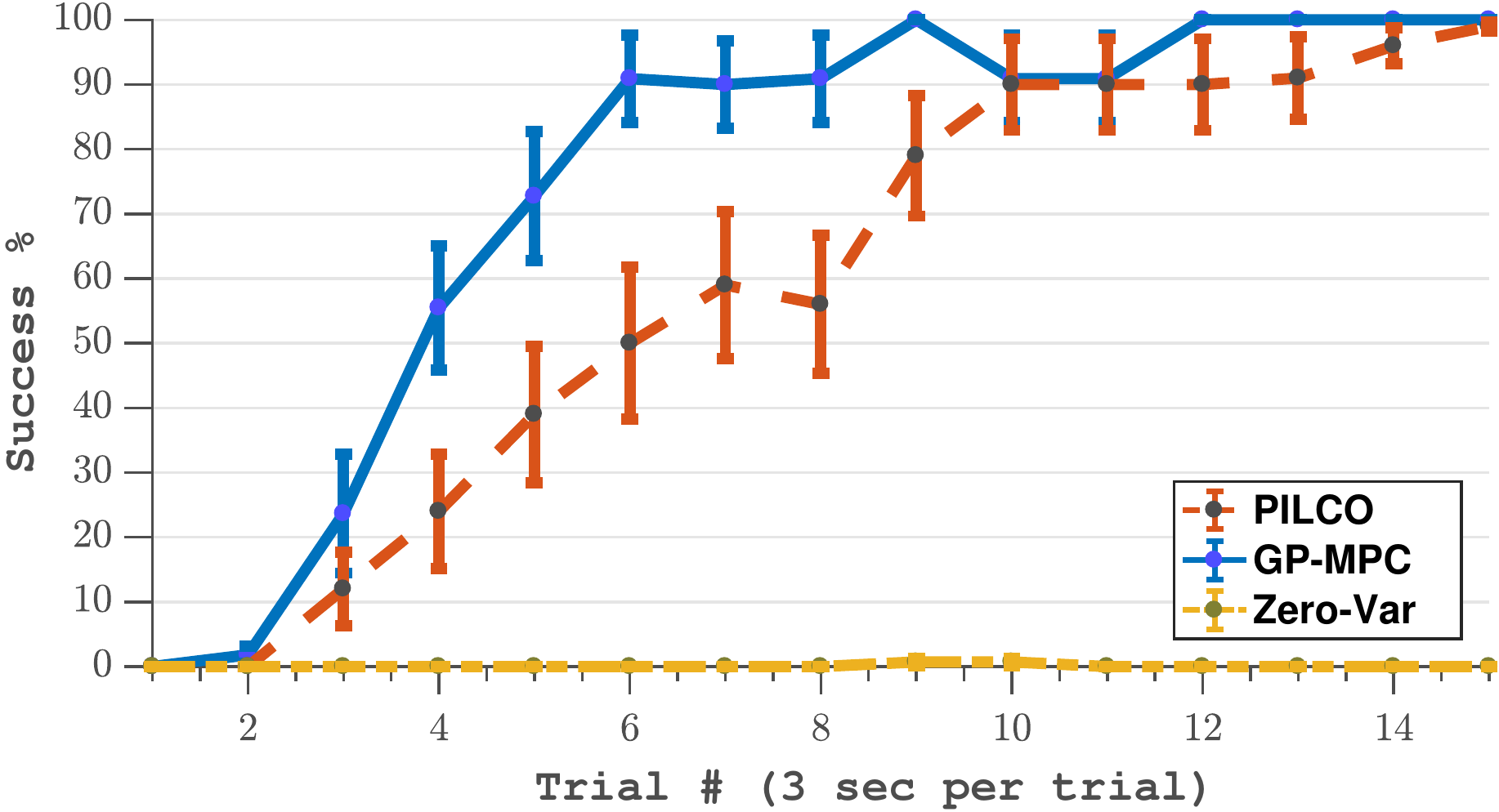}
    \label{fig:doublePend}
    }
	\caption{Performance of RL algorithms. Error bars represent the standard error. \subref{fig:cartpole} Cart-pole; \subref{fig:doublePend} Double pendulum. \modelname~(blue) consistently outperforms PILCO (red) and the zero-variance MPC approach (yellow) in terms of data efficiency. While the zero-variance MPC approach works well on the cart-pole task, it fails in the double-pendulum task. We attribute this to the inability to explore the state space sufficiently well.}
	\label{fig:results}
\end{figure}
\begin{figure}[tb]
	\centering
	\subfigure[Cart-pole with constraint.]{
		\includegraphics[height = 3.8cm]{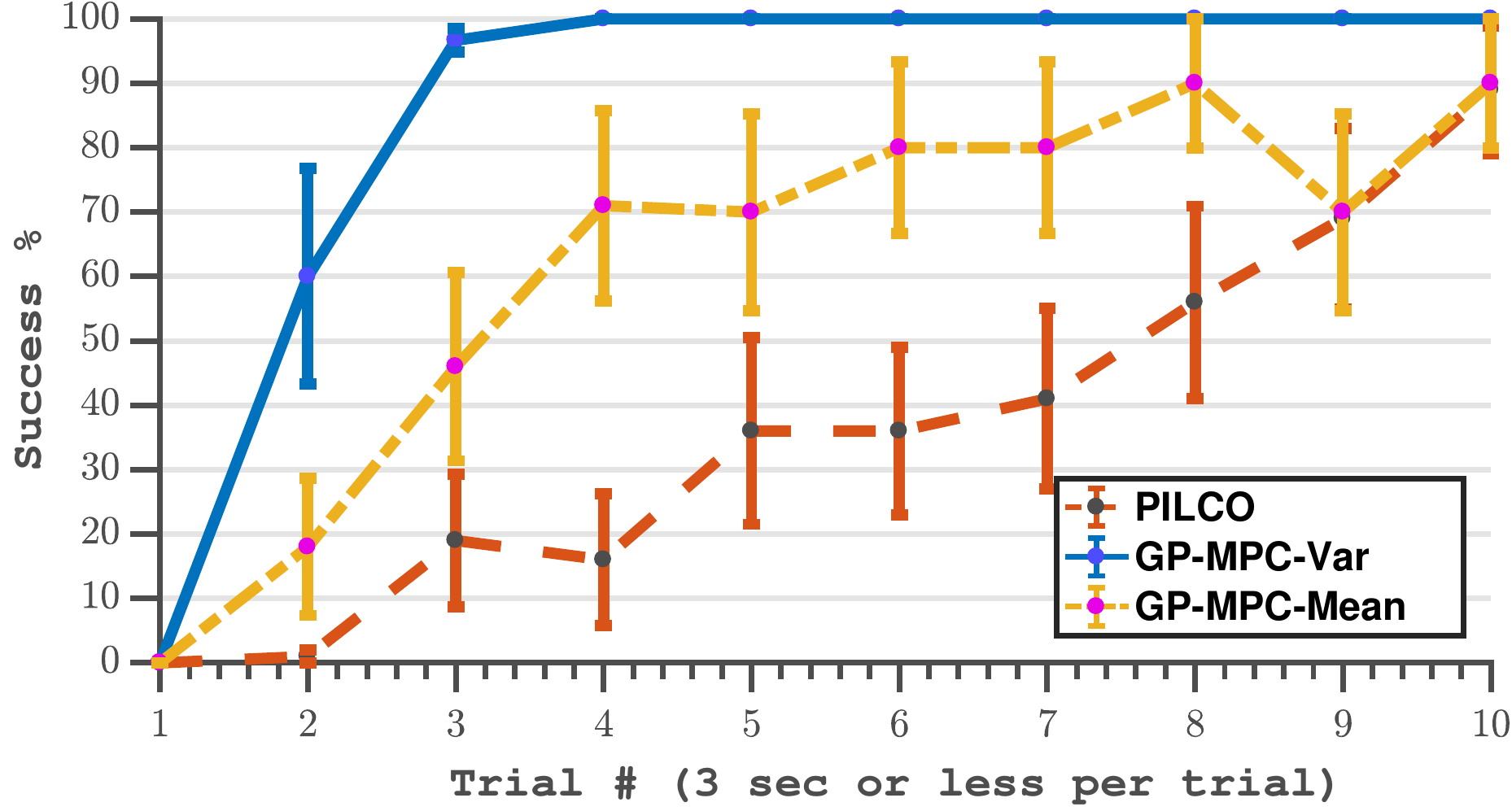}
		\label{fig:plot cp constraint}
	}
	\hfill
	\subfigure[Double pendulum with constraint.]{
		\includegraphics[height = 3.8cm]{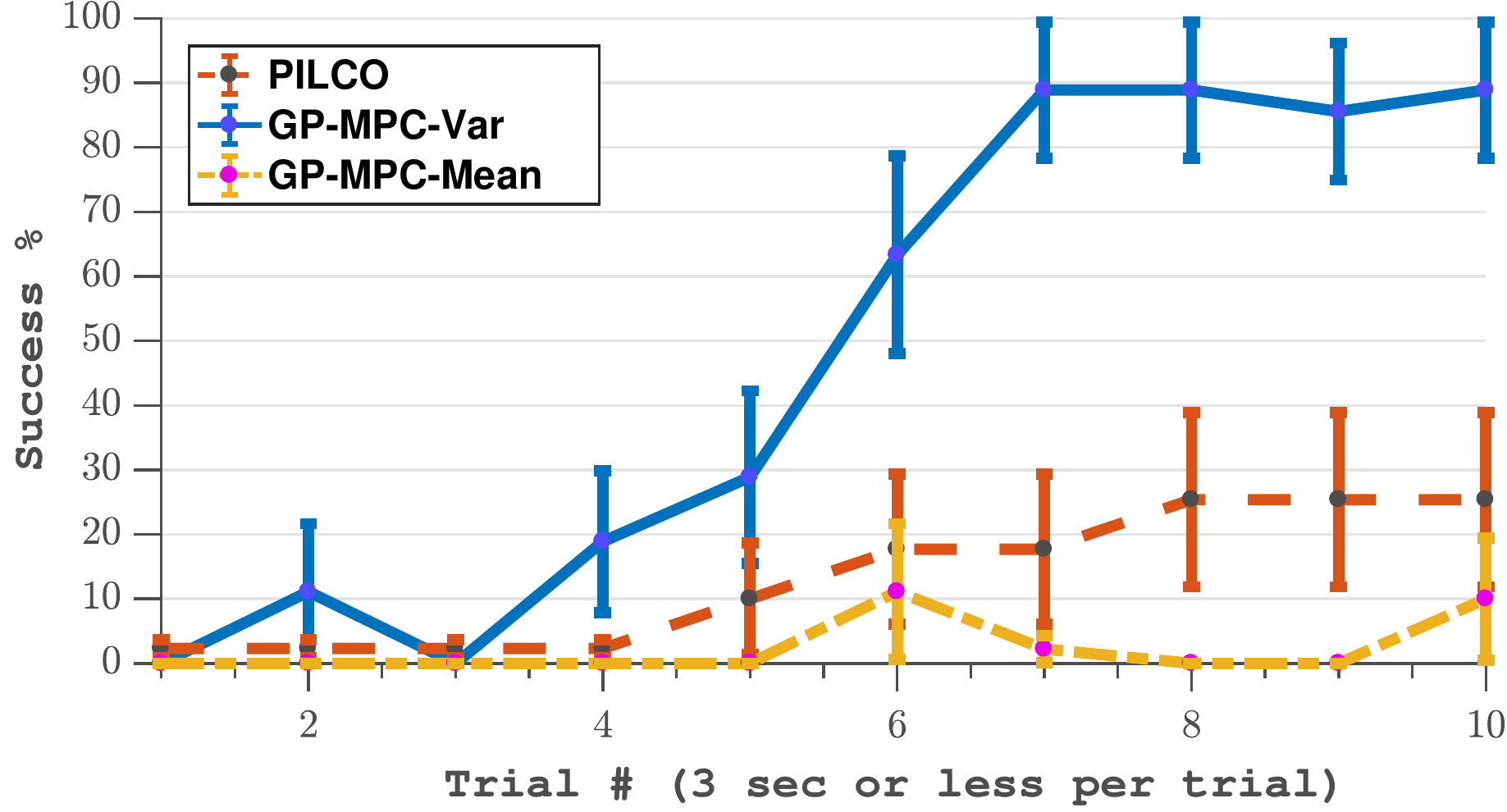}
		\label{fig:plot dp constraint}
	}
	\caption{Performance with state-space constraints. Error bars represent the standard error. \subref{fig:cp constraint} Cart-pole; \subref{fig:dp constraint} Double pendulum.\modelname~with chance constraints. GP-MPC-Var (blue) is the only method that is able to consistently solve the problem. Expected violations constraint GP-MPC-Mean (yellow) fails in cart-pole. PILCO (red) violates state constraints and struggles to complete the task.}
    \label{fig:plot state constraints}
\end{figure}
In both benchmark experiments (cart-pole and double pendulum), we use the exact saturating cost from~\citep{Deisenroth2015}, which penalises the Euclidean distance of the tip of the (outer) pendulum from the target position, i.e., we are in a setting in which PILCO performs very well. 


Fig.~\ref{fig:cartpole} shows that both our MPC-based controller (blue) and the zero-variance approach successfully\footnote{We define `success' if the pendulum tip is closer than $\unit[8]{cm}$ to the target position for ten consecutive time steps.} 
complete the task in fewer trials than the state-of-the-art PILCO method (red). From the repeated trials we see that \modelname~learns faster and more reliably than PILCO. In particular, \modelname~and the zero-variance approach can solve the cart-pole task with high probability (90\%) after three trials (9 seconds), where the first trial was random. PILCO needs two additional trials. The reason why the zero-variance approach (no model uncertainties) works in an MPC context but not within a policy search setting is that we include every observed state transition immediately in the GP dynamics model, which makes MPC fairly robust to model errors. It even allows for model-based RL with deterministic models in simple settings.

Fig.~\ref{fig:doublePend} highlights that our proposed \modelname~approach (yellow) requires on average only six trials ($\unit[18]{s}$) of experience to achieve a 90\% success rate\footnote{The tip of outer pendulum is closer than $\unit[22]{cm}$ to the target.}, including the first random trial. PILCO requires four additional trials, whereas the zero-variance MPC approach completely fails in this RL setting. The reason for this failure is that the deterministic predictions with a poor model in this complicated state space do not allow for sufficient exploration. We also observe that \modelname~is more robust to the variations amongst trials.

In both experiments, our proposed \modelname~requires only 60\% of PILCO's experience, such that we report an unprecedented learning speed for these benchmarks, even with settings for which PILCO performs very well.

We identify two key ingredients that are responsible for the success and learning speed of our approach: the ability to (1) immediately react to observed states by adjusting the long-term plan and (2) augment the training set of the GP model on the fly as soon as a new state transition is observed (hyper-parameters are not updated at every time step). These properties turn out to be crucial in the very early stages of learning when very little information is available. If we ignored the on-the-fly updates of the GP dynamics model, our approach would still successfully learn, although the learning efficiency would be slightly decreased. 

\subsection{State Constraints}

A scenario in which PILCO struggles is a setting with state space constraints. We modify the cart-pole and the double-pendulum tasks to such a setting. Both tasks are symmetric, and we impose state constraints in such a way that only one direction of the swing-up is feasible. For the cart-pole system, we place a wall near the target position of the cart, see Fig.~\ref{fig:cp constraint}; the double pendulum has a constraint on the angle of the inner pendulum, so that it only has a $340^\circ$ motion range, i.e., it cannot spin through, see Fig.~\ref{fig:dp constraint}. These constraints constitute linear constraints on the state.

We use a quadratic cost that penalises the Euclidean distance between the tip of the pendulum and the target. This, along with the `implicit' linearisation, makes the optimal control problem an `implicit' QP. If a rollout violates the state constraint we immediately abort that trial and move on to the next trial. We use the same experimental set-up as data efficiency experiments.  

The state constraints are implemented as \emph{expected violations}, i.e., $ \mathds{E} [\vec{x}_t] < \vec{x}_{\text{limit}} $ and \emph{chance constraints} $ p(\vec{x}_t< \vec{x}_{\text{limit}} ) \geq 0.95 $. 
Fig.~\ref{fig:cp constraint} shows that our MPC-based controller with chance constraint (blue) successfully completes the task with a small acceptable number of violations, see Table~\ref{tbl: violations}. The expected violation approach, which only considers the predicted mean (yellow) fails to complete the task due to repeated constraint violations. PILCO uses its saturating cost (there is little hope for learning with quadratic cost~\citep{Deisenroth2010}) and has partial success in completing the task, but it struggles, especially during initial trials due to repeated state violations. 

\begin{table}
\centering
\begin{tabular}{|c|c|c|}
	\hline 
	Experiment & Cart-pole & Double Pendulum \\ 
	\hline 
	PILCO & \red{16/100} & \red{23/100}  \\ 
	\hline 
	 GP-MPC-Mean & \red{21/100} & \red{26/100}  \\ 
	\hline 
	GP-MPC-Var & \blue{3/100} & \blue{11/100} \\
	\hline 
\end{tabular} 
\caption{State constraint violations. The number of trials that resulted in state constraint violation corresponding to the trial data shown in the the Fig.~\ref{fig:plot state constraints}.} 
\label{tbl: violations}
\end{table}

One of the key points we observe from the Table~\ref{tbl: violations} is that the incorporation of uncertainty into planning is again crucial for successful learning. If we use only predicted means to determine whether the constraint is violated, learning is not reliably `safe'. Incorporation of the predictive variance, however, results in significantly fewer constraint violations.

 \section{Conclusion and Discussion}
We proposed an algorithm for data-efficient RL that is based on probabilistic MPC with learned transition models using Gaussian processes. By exploiting Pontryagin's maximum principle our algorithm can naturally deal with state and control constraints. Key to this theoretical underpinning of a practical algorithm was the re-formulation of the optimal control problem with uncertainty propagation via moment matching into an deterministic optimal control problem. MPC allows the learned model to be updated immediately, which leads to an increased robustness with respect to model inaccuracies. We provided empirical evidence that our framework is not only theoretically sound, but also extremely data efficient, while being able to learn in settings with hard state constraints. 

One of the most critical components of our approach is the incorporation of model uncertainty into modelling and planning. In complex environments, model uncertainty drives targeted exploration. It additionally allows us to account for constraints in a risk-averse way, which is important in the early stages of learning.

\bibliographystyle{abbrv}


\begin{thebibliography}{10}

\bibitem{Bellman1957}
R.~E. Bellman.
\newblock {\em {Dynamic Programming}}.
\newblock Princeton University Press, Princeton, NJ, USA, 1957.

\bibitem{Bischoff2013}
B.~Bischoff, D.~Nguyen-Tuong, T.~Koller, H.~Markert, and A.~Knoll.
\newblock {Learning Throttle Valve Control Using Policy Search}.
\newblock In {\em Proceedings of the European Conference on Machine Learning
  and Knowledge Discovery in Databases}, 2013.

\bibitem{Bischoff2014}
B.~Bischoff, D.~Nguyen-Tuong, H.~van Hoof, A.~McHutchon, C.~E. Rasmussen,
  A.~Knoll, J.~Peters, and M.~P. Deisenroth.
\newblock {Policy Search for Learning Robot Control using Sparse Data}.
\newblock In {\em Proceedings of the International Conference on Robotics and
  Automation}, 2014.

\bibitem{Bock1984}
H.~G. Bock and K.~J. Plitt.
\newblock {A Multiple Shooting Algorithm for Direct Solution of Optimal Control
  Problems}.
\newblock In {\em Proceedings 9th IFAC World Congress Budapest}. Pergamon Press, 1984.

\bibitem{Boedecker2014}
J.~Boedecker, J.~T. Springenberg, J.~Wulfing, and M.~Riedmiller.
\newblock {Approximate Real-time Optimal Control based on Sparse Gaussian
  Process Models}.
\newblock In {\em Symposium on Adaptive Dynamic Programming and
  Reinforcement Learning}, 2014.

\bibitem{Clarke1990}
F.~H. Clarke.
\newblock {\em {Optimization and Non-Smooth Analysis}}.
\newblock Society for Industrial and Applied Mathematics, 1990.

\bibitem{Cutler2015}
M.~Cutler and J.~P. How.
\newblock {Efficient Reinforcement Learning for Robots using Informative
  Simulated Priors}.
\newblock In {\em Proceedings of the International Conference on Robotics and
  Automation}, 2015.
  
\bibitem{Deisenroth2010}
M.~P. Deisenroth
\newblock {\em Efficient Reinforcement Learning using Gaussian Processes}.
\newblock KIT Scientific Publishing, 2010.

\bibitem{Deisenroth2015}
M.~P. Deisenroth, D.~Fox, and C.~E. Rasmussen.
\newblock {Gaussian Processes for Data-Efficient Learning in Robotics and
  Control.}
\newblock {\em Transactions on Pattern Analysis and Machine Intelligence},
  37(2):408--23, 2015.

\bibitem{Deisenroth2011c}
M.~P. Deisenroth and C.~E. Rasmussen.
\newblock {PILCO: A Model-Based and Data-Efficient Approach to Policy Search}.
\newblock In {\em Proceedings of the International Conference on Machine
  Learning}, 2011.

\bibitem{Diehl2014}
M.~Diehl.
\newblock {\em {Lecture Notes on Optimal Control and Estimation}}.
\newblock 2014.

\bibitem{Girard2003}
A.~Girard, C.~E. Rasmussen, J.~Quinonero-Candela, and R.~Murray-Smith.
\newblock {Gaussian Process Priors with Uncertain Inputs-Application to
  Multiple-Step Ahead Time Series Forecasting}.
\newblock {\em Advances in Neural Information Processing Systems}, 2003.

\bibitem{Grancharova2008}
A.~Grancharova, J.~Kocijan, and T.~A. Johansen.
\newblock {Explicit Stochastic Predictive Control of Combustion Plants based on
  Gaussian Process Models}.
\newblock {\em Automatica}, 44(6):1621--1631, 2008.

\bibitem{Griewank1982}
A.~Griewank and P.~L. Toint.
\newblock {Partitioned Variable Metric Updates for Large Structured
  Optimization Problems}.
\newblock {\em Numerische Mathematik}, 39(1):119--137, 1982.

\bibitem{Grune2011}
L.~Gr{\"{u}}ne and J.~Pannek.
\newblock {Stability and Suboptimality Using Stabilizing Constraints}.
\newblock In {\em Nonlinear Model Predictive Control Theory and Algorithms},
  pages 87--112. Springer, 2011.

\bibitem{Hennig2011}
P.~Hennig.
\newblock {Optimal Reinforcement Learning for Gaussian Systems}.
\newblock In {\em Advances in Neural Information Processing
  Systems}, 2011.

\bibitem{Jaderberg2016}
M.~Jaderberg, V.~Mnih, W.~M. Czarnecki, T.~Schaul, J.~Z. Leibo, D.~Silver, and
  K.~Kavukcuoglu.
\newblock {Reinforcement Learning with Unsupervised Auxiliary Tasks}.
\newblock In {\em International Conference on Learning Representations}, 2016.

\bibitem{Klenske2016}
E.~D. Klenske, M.~N. Zeilinger, B.~Sch{\"{o}}lkopf, and P.~Hennig.
\newblock {Gaussian Process-Based Predictive Control for Periodic Error
  Correction}.
\newblock {\em Transactions on Control Systems Technology},
  24(1):110--121, 2016.

\bibitem{Kocijan2016}
J.~Kocijan.
\newblock {\em {Modelling and Control of Dynamic Systems Using Gaussian Process
  Models}}.
\newblock Advances in Industrial Control. Springer International Publishing, 2016.

\bibitem{Lee2017}
G.~Lee, S.~S. Srinivasa, and M.~T. Mason.
\newblock {GP-ILQG: Data-driven Robust Optimal Control for Uncertain Nonlinear
  Dynamical Systems}.
  \newblock{\em arXiv:1705.05344}
\newblock 2017.

\bibitem{MacKay1998}
D.~J.~C. MacKay.
\newblock {Introduction to Gaussian Processes}.
\newblock In  {\em Neural Networks and Machine Learning},
  volume 168, pages 133--165. Springer, Berlin, Germany, 1998.

\bibitem{Mayne2000}
D.~Q. Mayne, J.~B. Rawlings, C.~V. Rao, and P.~O. Scokaert.
\newblock {Constrained Model Predictive Control: Stability and Optimality}.
\newblock {\em Automatica}, 36(6):789--814, 2000.

\bibitem{Minka2001}
T.~P. Minka.
\newblock {\em {A Family of Algorithms for Approximate Bayesian Inference}}.
\newblock PhD thesis, Massachusetts Institute of Technology, Cambridge, MA,
  USA, 2001.

\bibitem{Minka2001a}
T.~P. Minka.
\newblock {Expectation Propagation for Approximate Bayesian Inference}.
\newblock In {\em Proceedings of the Conference on Uncertainty in Artificial Intelligence}, 2001.

\bibitem{Mnih2015}
V.~Mnih, K.~Kavukcuoglu, D.~Silver, A.~A. Rusu, J.~Veness, M.~G. Bellemare,
  A.~Graves, M.~Riedmiller, A.~K. Fidjeland, G.~Ostrovski, S.~Petersen,
  C.~Beattie, A.~Sadik, I.~Antonoglou, H.~King, D.~Kumaran, D.~Wierstra,
  S.~Legg, and D.~Hassabis.
\newblock {Human-Level Control through Deep Reinforcement Learning}.
\newblock {\em Nature}, 518(7540):529--533, 2015.

\bibitem{Naidu2003}
D.~S. D.~S. Naidu and R.~C. {Naidu, Subbaram/Dorf}.
\newblock {\em {Optimal Control Systems}}.
\newblock CRC Press, 2003.

\bibitem{Ng2000}
A.~Y. Ng and M.~I. Jordan.
\newblock {PEGASUS: A Policy Search Method for Large MDPs and POMDPs.}
\newblock {\em Proceedings of the Conference on Uncertainty in Artificial Intelligence}, 2000.

\bibitem{Nghiem2017}
T.~X. Nghiem and C.~N. Jones.
\newblock {Data-driven Demand Response Modeling and Control of Buildings with
  Gaussian Processes}.
\newblock In {\em Proceedings of the American Control Conference}, 2017.

\bibitem{Nocedal2006}
J.~Nocedal and S.~J. Wright.
\newblock {\em {Numerical Optimization}}.
\newblock Springer, 2006.

\bibitem{Opper2001}
M.~Opper and O.~Winther.
\newblock {Tractable Approximations for Probabilistic Models: The Adaptive TAP
  Mean Field Approach}.
\newblock {\em Physical Review Letters}, 86(17):5, 2001.

\bibitem{Ostafew2015}
C.~Ostafew, A.~Schoellig, T.~Barfoot, and J.~Collier.
\newblock {Learning-based Nonlinear Model Predictive Control to Improve
  Vision-based Mobile Robot Path Tracking}.
\newblock {\em Journal of Field Robotics}, 33(1):133--152, 2015.

\bibitem{Ostafew2016}
C.~Ostafew, A.~Schoellig, T.~Barfoot.
\newblock {Robust Constrained Learning-based NMPC enabling reliable mobile robot path tracking}.
\newblock {\em 
The International Journal of Robotics Research}, 35(13):1547-1563, 2016.

\bibitem{Pan2014}
Y.~Pan and E.~Theodorou.
\newblock {Probabilistic Differential Dynamic Programming}.
\newblock {\em Advances in Neural Information Processing Systems}, 2014.

\bibitem{Pan2015}
Y.~Pan, E.~Theodorou, and M.~Kontitsis.
\newblock {Sample Efficient Path Integral Control under Uncertainty}.
\newblock{\em Advances in Neural Information Processing Systems}, 2015.

\bibitem{Polydoros2017}
A.~S. Polydoros and L.~Nalpantidis.
\newblock {Survey of Model-Based Reinforcement Learning: Applications on
  Robotics}.
\newblock {\em Journal of Intelligent and Robotic Systems}, 86(2):153--173,
  2017.

\bibitem{Pontryagin1962}
L.~S. Pontryagin, E.~F. Mishchenko, V.~G. Boltyanskii, and R.~V. Gamkrelidze.
\newblock {\em {The Mathematical Theory of Optimal Processes}}.
\newblock Wiley, 1962.

\bibitem{Quinonero-Candela2005a}
J.~Qui{\~{n}}onero-Candela and
  C.~E. Rasmussen.
\newblock {A Unifying View of Sparse Approximate Gaussian Process Regression}.
\newblock {\em Journal of Machine Learning Research}, 6(2):1939--1960,
  2005.

\bibitem{Raiko2009}
T.~Raiko and M.~Tornio.
\newblock {Variational Bayesian Learning of Nonlinear Hidden State-Space Models
  for Model Predictive Control}.
\newblock {\em Neurocomputing}, 72(16--18):3702--3712, 2009.

\bibitem{Rasmussen2006}
C.~E. Rasmussen and C.~K.~I. Williams.
\newblock {\em {Gaussian Processes for Machine Learning}}.
\newblock The MIT Press, Cambridge,
  MA, USA, 2006.

\bibitem{Rawlik2012}
K.~Rawlik, M.~Toussaint, and S.~Vijayakumar.
\newblock {On Stochastic Optimal Control and Reinforcement Learning by
  Approximate Inference}.
\newblock In {\em Proceedings of Robotics: Science and Systems}, 2012.

\bibitem{Schattler2012}
H.~Sch{\"{a}}ttler and U.~Ledzewicz.
\newblock {\em {Geometric Optimal Control: Theory, Methods and Examples}},
  volume~53.
\newblock 2012.

\bibitem{Schneider1997}
J.~G. Schneider.
\newblock {Exploiting Model Uncertainty Estimates for Safe Dynamic Control
  Learning}.
\newblock In {\em Advances in Neural Information Processing Systems}. 1997.

\bibitem{Silver2016}
D.~Silver, A.~Huang, C.~J. Maddison, A.~Guez, L.~Sifre, G.~van~den Driessche,
  J.~Schrittwieser, I.~Antonoglou, V.~Panneershelvam, M.~Lanctot, S.~Dieleman,
  D.~Grewe, J.~Nham, N.~Kalchbrenner, I.~Sutskever, T.~Lillicrap, M.~Leach,
  K.~Kavukcuoglu, T.~Graepel, and D.~Hassabis.
\newblock {Mastering the Game of Go with Deep Neural Networks and Tree Search}.
\newblock {\em Nature}, 529(7587), 2016.

\bibitem{Sutton1998}
R.~S. Sutton and A.~G. Barto.
\newblock {\em {Reinforcement Learning: An Introduction}}.
\newblock The MIT Press, Cambridge,  MA, USA, 1998.

\bibitem{Todorov2009}
E.~Todorov.
\newblock {Efficient Computation of Optimal Actions.}
\newblock {\em Proceedings of the National Academy of Sciences of the United
  States of America}, 106(28):11478--83, 2009.

\bibitem{Todorov2005}
E.~Todorov and {Weiwei Li}.
\newblock {A Generalized Iterative LQG Method for Locally-Optimal Feedback
  Control of Constrained Nonlinear Stochastic Systems}.
\newblock In {\em Proceedings of the  American Control Conference.},  2005.

\bibitem{Toussaint2009}
M.~Toussaint.
\newblock {Robot Trajectory Optimization using Approximate Inference}.
\newblock In {\em Proceedings of the International Conference on Machine
  Learning}, 2009.

\bibitem{Yahya2016}
A.~Yahya, A.~Li, M.~Kalakrishnan, Y.~Chebotar, and S.~Levine.
\newblock {Collective Robot Reinforcement Learning with Distributed
  Asynchronous Guided Policy Search}.
\newblock {\em arXiv:1610.00673}, 2016.

\end{thebibliography}

\begin{thebibliography}{10}
\bibitem{Deisenroth2015}
M.~P. Deisenroth, D.~Fox, and C.~E. Rasmussen.
\newblock {Gaussian Processes for Data-Efficient Learning in Robotics and
  Control.}
\newblock {\em Transactions on Pattern Analysis and Machine Intelligence},
  37(2):408--23, 2015.
\end{thebibliography}

\section*{Appendix}

\subsection*{Lipschitz Continuity}
\begin{lemma}
\label{lemma:fmm L-continuous appendix}
	The moment matching mapping $ f_{MM} $ is Lipschitz continuous for controls defined over a compact set $\mathcal{U}$. 
\end{lemma}

\textbf{Proof: }
Lipschitz continuity requires that the gradient  $ {\partial f_{MM}}/{\partial \vec u_t} $ is bounded.  The gradient is 
\begin{align}
\frac{\partial f_{MM}}{\partial \vec u_t} =
\frac{\partial \vec z_{t+1}}{\partial\vec u_t} = \left[ \frac{\partial \vec \mu_{t+1}}{\partial \vec u_t} , \frac{\partial \vec \Sigma_{t+1}}{\partial \vec u_t} \right]\,.
\end{align}
The derivatives $  \left[ \frac{\partial\vec \mu_{t+1}}{\partial \vec u_t}, \frac{\partial \vec \Sigma_{t+1}}{\partial\vec  u_t} \right]$ can be computed analytically~\citep{Deisenroth2015}.

We first show that the derivative $\partial \vec\mu_{t+1}/\partial\vec u_t$ is bounded. Defining $\vec\beta_d:= (\mat K_d + \sigma_{f_d}^2\mat I)\inv\vec y_d$, from~\citep{Deisenroth2015}, we obtain for all state dimensions $d=1,\dotsc, D$
\begin{align}
   \mu_{t+1}^d &=  \sum\nolimits_{i=1}^N \beta_{d_i}q_{d_i}\,, \\
   q_{d_i} &= \sigma_{f_d}^2 |\mat I
  +\mat L_d\inv\aug{\mat\Sigma}_{t}|^{-\tfrac{1}{2}} \, \times \\
  &\quad \exp\big(-\tfrac{1}{2}
  (\aug{\vec x}_i -\aug{\vec\mu}_{t})\T(\mat L_d +
  \aug{\mat\Sigma}_{t})\inv (\aug{\vec x}_i
  -\aug{\vec\mu}_{t})\big)\,,
  \label{eq:def q_i}
\end{align}
where $N$ is the size of the training set of the dynamics GP and $\aug{\vec x}_i$ the $i$th training input. The corresponding gradient w.r.t. $\vec u_t$ is given by the last $F$ elements of
\begin{align}
  \frac{\partial\mu_{t+1}^d}{\partial\aug{\vec\mu}_{t}} & =
  \sum\nolimits_{i=1}^N \beta_{d_i}\frac{\partial
    q_{d_i}}{\partial\aug{\vec\mu}_{t}} \\ & = \sum\nolimits_{i=1}^N\beta_{d_i} q_{d_i}(\aug{\vec x}_i -
  \aug{\vec\mu}_{t})\T(\aug{\mat\Sigma}_{t} +
  \mat L_d)\inv \in\R^{1\times (D+F)}
\label{eq:dmu/dmu}
\end{align}


Let us examine the individual terms in the sum on the rhs in~\eqref{eq:dmu/dmu}: 
For a given trained GP $ \|\vec \beta_d\|<\infty$ is constant. The definition of $q_{d_i}$ in~\eqref{eq:def q_i} contains an exponentiated negative quadratic term, which is bounded between $[0,1]$. Since $\mat I +  \mat L_d\inv \aug{\mat\Sigma}_t$ is positive definite, the inverse determinant is defined and bounded. Finally, $\sigma_{f_d}^2 <\infty$, which makes $q_{d_i}<\infty$. The remaining term in~\eqref{eq:dmu/dmu} is a vector-matrix product. The matrix is regular and its inverse exists and is bounded (and constant as a function of $\vec u_t$. Since $\vec u_t\in \mathcal U$ where $\mathcal U$ is compact, we can also conclude that the vector difference in~\eqref{eq:dmu/dmu} is finite, which overall proves that $ f_{MM} $ is (locally) Lipschitz continuous and Lemma~\ref{lemma:fmm L-continuous appendix}.

\newpage
\subsection*{Sequential Quadratic Programming}
We can use SQP for solving non-linear optimization
problems (NLP) of the form,
\[
\begin{array}{rl}
\min\limits_{ u} & f(\vec x) \\
\mbox{s.t.} & b(\vec x) \ge 0 \\
  & \vec c(\vec x) = 0.
\end{array}\]
The Lagrangian $\mathcal{L}$ associated with the NLP is  
\begin{align}
\label{defn:Lagrangian}
\mathcal{L}(\vec x,\vec \lambda,\vec \sigma) = f(\vec x) - \vec \sigma^T b(\vec x) - \vec \lambda^T c(\vec x)
\end{align}

where, $\vec \lambda$ and $\vec \sigma$ are Lagrange multipliers.
Sequential Quadratic Programming (SQP) forms a quadratic (Taylor) approximation of the objective and linear approximation of constraints at each iteration $k$
\begin{align}
\begin{array}{rl} \min\limits_{d} & f(\vec x_k) + \nabla f(\vec x_k)^T \vec d + \tfrac{1}{2} \vec d^T \mat \nabla_{xx}^2 \mathcal{L}(\vec x,\vec \lambda,\vec \sigma) \vec d \\
\mathrm{s.t.} & b(\vec x_k) + \nabla b(\vec x_k)^T \vec d \ge 0 \\
  & c(\vec x_k) + \nabla c(\vec x_k)^T \vec d = 0. \end{array}
\end{align}

  The Lagrange multipliers $\vec \lambda$ associated with the equality constraint are same as the ones defined in the control Hamiltonian $\Hamiltonian$~\ref{defn:hamiltonian}. The Hessian matrix $\mat \nabla_{xx}^2$ can be computed by exploiting the block diagonal structure introduced by the Hamiltonian~\citep{Griewank1982, Bock1984}.


 \subsection{Moment Matching Approximation~\cite{Deisenroth2015}}
\label{sec:moment matching}
Following the law of iterated expectations, for target dimensions
$a=1,\dotsc,D,$ we obtain the \emph{predictive mean}
\begin{align}
  \mu_{t}^a&=\E_{\tilde{\vec x}_{t-1}}[\E_{f_a}[f_a(\tilde{\vec
    x}_{t-1})|\tilde{\vec x}_{t-1}]]= \E_{\tilde{\vec
      x}_{t-1}}[m_{f_a}(\tilde{\vec x}_{t-1})]\nonumber\\ 
  &= \int m_{f_a}(\tilde{\vec x}_{t-1})\gaussx{\tilde{\vec
      x}_{t-1}}{\tilde{\vec\mu}_{t-1}}{\tilde{\mat\Sigma}_{t-1}}\d
  \tilde{\vec x}_{t-1}\nonumber\\
  &=\vec{\beta}_a\T\vec q_a\,,
  \label{eq: pred. mean uncertain input}\\
  \vec\beta_a &= (\mat K_a + \sigma_{w_a}^2)\inv\vec y_a
\label{eq:beta}
\end{align}
with $\vec q_a = [q_{a_1}, \ldots, q_{a_n}]\T$. The entries of $\vec
q_a \in\R^n$ are computed using standard results from multiplying and
integrating over Gaussians and are given by
\begin{align}
  q_{a_i} &= \int k_a(\tilde{\vec x}_i,\tilde{\vec
    x}_{t-1})\gaussx{\tilde{\vec
      x}_{t-1}}{\tilde{\vec\mu}_{t-1}}{\tilde{\mat\Sigma}_{t-1}}
  \d\tilde{\vec x}_{t-1}\label{eq:q_i}\\
  & = \sigma_{f_a}^2|\tilde{\mat\Sigma}_{t-1}\mat\Lambda_a\inv + \mat
  I|^{-\frac{1}{2}}\exp\big(-\tfrac{1}{2}\vec\nu_i\T(\tilde{\mat\Sigma}_{t-1}
  + \mat\Lambda_a)\inv\vec\nu_i\big) \nonumber\,,
\end{align}
where  we define
\begin{align}
  \vec\nu_i\coloneqq (\tilde{\vec x}_i - \tilde{\vec\mu}_{t-1})
\label{eq:nu_i}
\end{align}
is the difference between the training input $\tilde{\vec x}_i$ and
the mean of the test input distribution $\prob(\vec x_{t-1},\vec
u_{t-1})$.

Computing the \emph{predictive covariance matrix}
$\mat\Sigma_{t}\in\R^{D\times D}$ requires us to distinguish
between diagonal elements and off-diagonal elements: Using the law of
total \mbox{(co-)}variance, we obtain for target dimensions
$a,b=1,\dotsc,D$
\begin{align}
  \sigma_{aa}^2 &\!=\! \E_{\tilde{\vec x}_{t-1}}\big[\var_f
  [x_t^a|\tilde{\vec x}_{t-1}]\big]\!+\! \E_{f,\tilde{\vec
    x}_{t-1}}[({\vec x_t^a})^2]\!-\!(\mu_{t}^a)^2
\label{eq:diagonal entry def.}\,,\\
\sigma_{ab}^2 &\!=\! \E_{f,\tilde{\vec
    x}_{t-1}}[x_t^ax_t^b]\!-\!\mu_{t}^a\mu_{t}^b\,,\quad
a\neq b\,,
\label{eq:off-diagonal entry def}
\end{align}
respectively, where $\mu_t^a$ is known from (\ref{eq:
  pred. mean uncertain input}). The off-diagonal terms do not contain
the additional term $\E_{\tilde{\vec x}_{t-1}}[\cov_f
[x_t^a,x_t^b|\tilde{\vec x}_{t-1}]]$ because of the conditional
independence assumption of the GP models. Different target dimensions
do not covary for given $\tilde{\vec x}_{t-1}$.

We start the computation of the covariance matrix with the terms that
are common to both the diagonal and the off-diagonal entries: With
$\prob(\tilde{\vec x}_{t-1}) = \gaussx{\tilde{\vec
    x}_{t-1}}{\tilde{\vec\mu}_{t-1}}{\tilde{\mat\Sigma}_{t-1}}$ and
the law of iterated expectations, we obtain
\begin{align}
  \E_{f,\tilde{\vec
      x}_{t-1}}&[x_t^a,x_t^b]=\E_{\tilde{\vec
      x}_{t-1}}\big[\E_f[x_t^a|\tilde{\vec
    x}_{t-1}]\,\E_f[x_t^b|\tilde{\vec x}_{t-1}]\big]\nonumber\\
  &=\!\int\! m_f^a(\tilde{\vec
    x}_{t-1}) m_f^b(\tilde{\vec x}_{t-1})\prob(\tilde{\vec
    x}_{t-1})\d\tilde{\vec x}_{t-1}
\end{align}
because of the conditional independence of $x_t^a$ and $x_t^b$
given $\tilde{\vec x}_{t-1}$. Using the definition of the mean
function, we obtain
\begin{align}
  &\E_{f,\tilde{\vec x}_{t-1}}[x_t^ax_t^b] =\vec\beta_a\T\mat
  Q\vec\beta_b\,,
\label{eq:off-diagonal intermediate result}
\\
&\mat Q\coloneqq\int k_a(\tilde{\vec x}_{t-1},\mat X)\T \, k_b(\tilde{\vec
  x}_{t-1},\mat X)\prob(\tilde{\vec x}_{t-1})\d\tilde{\vec x}_{t-1}\,.
\label{eq:definition Q matrix as an integral}
\end{align}
Using standard results from Gaussian multiplications and integration,
we obtain the entries $Q_{ij}$ of $\mat Q\in\R^{n\times n}$
\begin{align}
  Q_{ij}
&=\frac{k_a(\tilde{\vec x}_i,\tilde{\vec\mu}_{t-1})k_b(\tilde{\vec
    x}_j, \tilde{\vec\mu}_{t-1})}{\sqrt{|\mat R|}}\exp\big(\tfrac{1}{2}\vec
  z_{ij}\T\mat T\inv \vec
  z_{ij}\big)
\label{eq:Q-matrix entries}
\end{align}
where we define
\begin{align*}
  \mat R&\coloneqq
  \tilde{\mat\Sigma}_{t-1}(\mat\Lambda_a\inv+\mat\Lambda_b\inv)+\mat
  I\,,\quad \mat T \coloneqq \mat\Lambda_a\inv + \mat\Lambda_b\inv +\tilde{\mat\Sigma}_{t-1}\inv\,,\\
  \vec
  z_{ij}&\coloneqq\mat\Lambda_a\inv\vec\nu_i+\mat\Lambda_b\inv\vec\nu_j\,,
\end{align*}
with $\vec \nu_i$ taken from (\ref{eq:nu_i}).  Hence, the
off-diagonal entries of $\mat\Sigma_t$ are fully determined by~(\ref{eq: pred. mean uncertain input})--(\ref{eq:nu_i}),
(\ref{eq:off-diagonal entry def}),~(\ref{eq:off-diagonal intermediate
  result}),~(\ref{eq:definition Q matrix as an integral}),
and~(\ref{eq:Q-matrix entries}).

\bibliographystyle{abbrv}


\end{document}